\documentclass[%
 reprint,
superscriptaddress,
showkeys,
 amsmath,amssymb,
 aps,
prl,
]{revtex4-1}

\usepackage{graphicx}
\usepackage[caption=false]{subfig} 

\usepackage{dcolumn}
\usepackage{bm}

\begin{document}

\preprint{APS/123-QED}
\title{The escape problem for active particles confined to a disc} 

\author{Kristian St\o{}levik Olsen}
\affiliation{PoreLab, Department of Physics, University of Oslo, Blindern, 0316 Oslo, Norway\\}

\author{Luiza Angheluta}
\affiliation{PoreLab, Department of Physics, University of Oslo, Blindern, 0316 Oslo, Norway\\}

\author{Eirik Grude Flekk\o{}y}
\affiliation{PoreLab, Department of Physics, University of Oslo, Blindern, 0316 Oslo, Norway\\}

\date{\today}

\begin{abstract}
We study the escape problem for interacting, self-propelled particles confined to a disc, where particles can exit through one open slot on the circumference. Within a minimal 2D Vicsek model, we numerically study the statistics of escape events when the self-propelled particles can be in a flocking state. We show that while an exponential survival probability is characteristic for non-interaction self-propelled particles at all times, the interacting particles have an initial exponential phase crossing over to a sub-exponential late-time behavior. We propose a new phenomenological model based on non-stationary Poisson processes which includes the Allee effect to explain this sub-exponential trend and perform numerical simulations for various noise intensities. 
\end{abstract}

\pacs{Valid PACS appear here} 

\maketitle


A common trait for many soft active matter systems, formed by the self-propelled (active) individuals, is their ability to self-organize into complex flowing states that arise due to many-body interactions and an energy input on the particle level \cite{marchetti2013hydrodynamics,bar2020self}. A wide range of systems live under the umbrella of active matter including biological microswimmers \cite{dombrowski2004self,riedel2005self}, Janus particles \cite{walther2008janus,jiang2010active,bickel2014polarization} and vibrated granular rods \cite{kumar2014flocking, kudrolli2008swarming}, and most of these systems are embedded into an environment or a spatial confinement which can alter the open-space particle dynamics \cite{crowdedactive}. Recently, experiments and simulations have shown that the interactions between the self-propelled particles or interactions with obstacles and boundaries give rise to interesting behaviours like particle migration towards walls \cite{volpe2014simulation,yang2014aggregation}, separation in systems with more than one type of active particles \cite{mijalkov2013sorting}, as well as trapping \cite{kumar2019trapping}. The role of confinement of active particles in undoubtedly fundamental for in realistic systems especially for biological matter and biotechnology \cite{palagi2018bioinspired,gompper20202020}. However, it is also one of the least understood and open topic in current active matter research. The confinement introduces a length scale into the problem, which interacts with the many other length-scales that are already present in active matter, changing the emergent pattern formations in the flocking states.

\begin{figure}[ht]
   \centering
   \includegraphics[width = 6.5 cm]{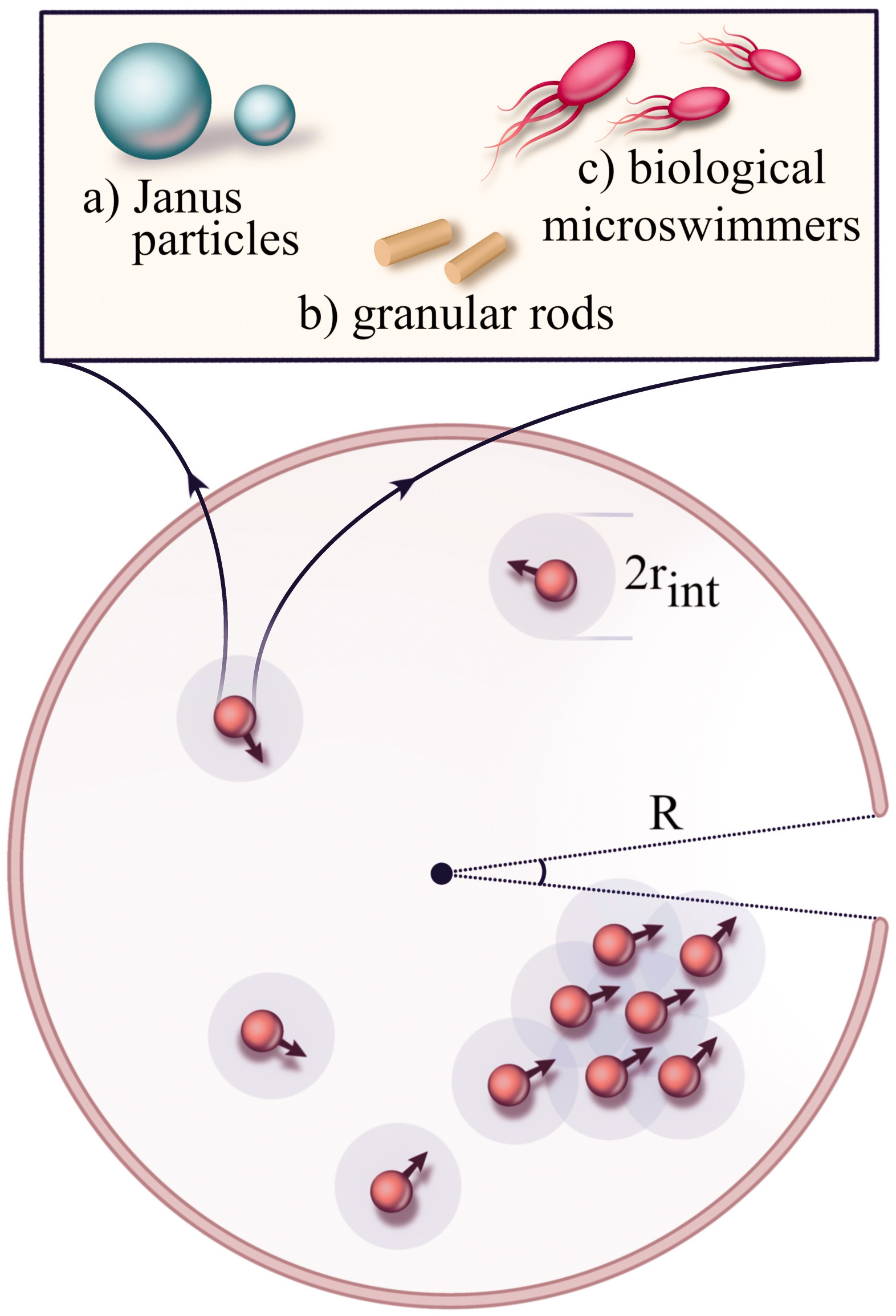}
    \caption{A sketch of the system considered in this paper. The particles may represent self-propelled agents like Janus particles, vibrated granular rods or biological microswimmers, modeled for simplicity as spherically symmetric polar particles with a small volume exclusion radius $r_{ve} = 1$ in units of the particle step length. The angular opening of the escape window is fixed to $2\pi/18$. The particles interact through a Vicsek-type alignment interaction with range $r_\text{int} = 5$, enabling collective escape events. }
    \label{fig:sketch}
\end{figure}

The narrow escape problem is a classical problem in statistical physics, where particles move inside a bounded 2D domain with a small part of the boundary being absorbing. The type of escape process is determined by the particle dynamics inside the domain, and various behaviors have been studied in the past. The classic example is that of Brownian motion which results in an exponential decay on the number of particles, or equivalently, the survival probability \cite{schuss2007narrow}. In recent times, the narrow escape problem has gained renewed interest due to its relevance in biological processes, where the absorbing window may for example represent a small patch of a cellular membrane where receptors are located, and the diffusing particle represents an ion \cite{schuss2007narrow,singer2008narrow}. An exponential decay is also found in chaotic billiard systems, while deterministic billiards give rise to a $1/t$ decay in particle number \cite{bauer1990decay}. The survival probability in a 1D setting has also recently been studied in a run-and-tumble model of bacterial motion \cite{mori2020universal}.

In this Letter, we study the problem of interacting active particles confined to a disc with a small opening through which they may escape, as depicted in Fig.(\ref{fig:sketch}). In the high-noise, weak-interaction limit the problem is similar to that of the Brownian escape problem in the sense that interactions are negligible, while in the opposite regime, we expect collective effects to alter the escape process leaving the particle number decay non-exponentially. It is the low-noise regime that is of primary interest in this Letter. We perform numerical simulations for both interacting and non-interacting self-propelled particles, and study the survival probability and escape time distribution. The simulations reveal a sub-exponential decay at late times whose origin we are able to explain within a minimal Poisson process with a non-stationary rate $\lambda$. A model with a density dependent rate inspired by models in population ecology is proposed, which reproduces the obtained event statistics.  

\begin{figure*}[t]
    \centering
    \includegraphics[width = \textwidth]{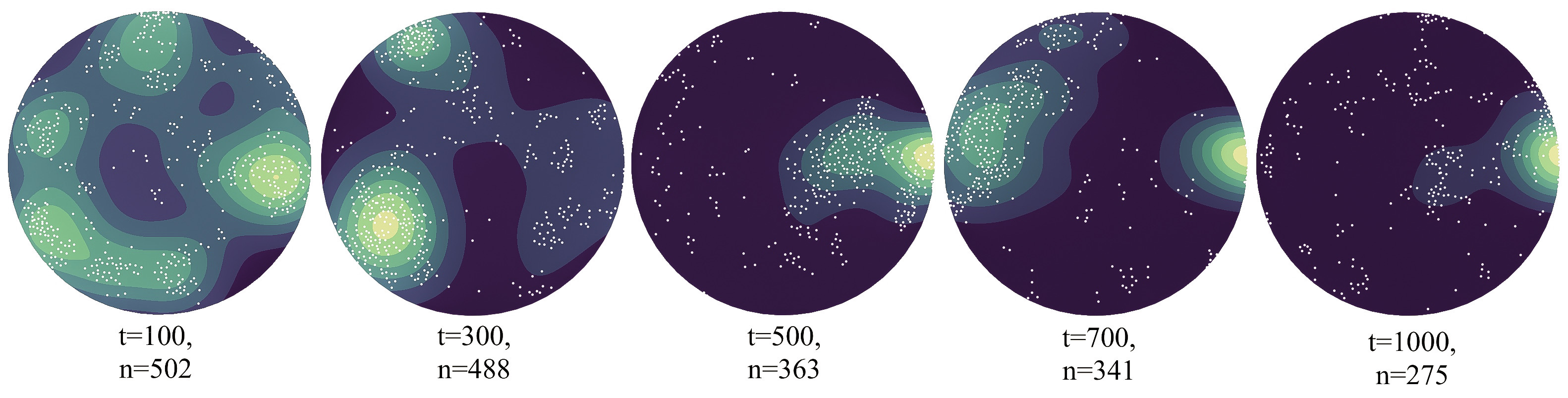}
    \caption{Snapshots of a simulation based on Eq. (\ref{eq:vic2})-(\ref{eq:vic3}) with active particles confined to a disc with an absorbing window with opening angle $2\pi/18$ centered at $(R,0)$. Background color map shows relative particle density for all the particles (absorbed and non-absorbed), bright implying high density. We see that as time progresses the density at the absorbing window increases as particles accumulate. System parameters: $\eta = 0.2, R = 100, r_\text{int} = 6, r_\text{ve} =2, N_0 = 2
   ^9$.}
    \label{fig:snaps}
\end{figure*}

In general terms, we consider active particles moving in a 2D bounded domain $\Omega$ with boundary $\partial \Omega = \partial \Omega_r \cup \partial \Omega_a$ where the subscripts denote the reflective and absorbing parts of the boundary respectively. The particles have a density $\rho(\vec{x},t)$, assumed to be normalized to unity at $t=0$, which follows a continuity equation. From this probability density the survival probability is defined as 
\begin{equation}\label{eq:Surv}
    S(t) = \int_\Omega d\Omega \rho(\vec{x},t).
\end{equation}
The first hitting time (FHT), in this case also the escape time, is the time $T_1$ at which a particle escapes the domain. The distribution of first hitting times $H(t)$ is closely related to the survival probability, namely
$$ S(t) = \int_t^\infty d s H(s),$$
which simply states that the probability of survival up to time $t$ is equivalent to the FHT being larger that $t$. This implies for the FHT distribution that 
\begin{equation}\label{eq:FHTdef}
H(t) = - \partial_t S(t).
\end{equation}
We see that the distribution of escape times can be interpreted as the probability flux out of the system.

Vicsek models are undoubtfully the archetypal numerical models for collective swarming and flocking effects \cite{vicsek1995novel,ginelli2016physics}. The particles are self-propelled with velocity $\dot {\vec{x}}_i = v_0 \hat{P}_i(\theta)$ where the polarization vector $\hat{P}_i(\theta)$ for particle $i$ is updated according to the standard Vicsek rule \cite{chate2008modeling}
\begin{align}
    & {\hat{P}}_i(t+1) = D_\eta \left(\frac{ \overline{v}_i }{||\overline{v}_i ||} \right), \label{eq:vic2}  \\
    &\overline{v}_i = \sum_{k :||\vec{x}_i-\vec{x}_k|| \leq r_\text{int}} \vec{v}_k, \label{eq:vic3}
\end{align}
 where  $D_{\eta}$ is a rotation matrix rotating a vector by a random uniformly chosen angle in $(-\eta \pi, \eta \pi)$. The parameter $\eta \in (0,1)$ determines the noise in the system. The velocity $\overline{v}_i $ in Eq. (\ref{eq:vic3}) is the average velocity of the neighboring particles of $i$, representing the velocity with which particle $i$ tries to align. The alignment interaction has a range $r_\text{int}$. Note that the velocity of particle $i$ itself is included in this sum leading to $\overline{v}_i$, so that in the non-interacting limit $r_\text{int}\to 0$ the particle moves according to a very simple stochastic model governed only by the parameter $\eta$ and the self-propulsion speed $v_0$, which we here set to unity without loss of generality.
 
  Eq. (\ref{eq:vic2}) and (\ref{eq:vic3}) must be supplemented with additional information when boundaries or obstacles are present. In the current case, the reflecting boundary of the disc can be simply dealt with by letting the director $\hat P_i$ be reflected about the tangent to the circle at the point of impact. A small volume exclusion  interaction with range $r_{\text{ve}}$ is numerically included by moving particles a step length apart is the direction separating them should they come to close to each other. This is necessary in confined spaces since a flock colliding with a wall would otherwise tend to collapse the particle density. Fig. (\ref{fig:snaps}) shows an example of the dynamics produced by this model. 
  
On the hydrodynamic scale, let us assume that the phase space density of an active particle $\Psi(\vec{x},\theta,t)$ satisfies a Boltzmann-type mean field equation $\mathcal{D}_t \Psi = Q[\Psi]$ where the total time derivative includes the self-propulsion term and takes the form $\mathcal{D}_t = \partial_t + v_0 \hat{P}(\theta) \cdot \nabla_x$. This must of course be supplemented with appropriate boundary conditions for the reflective and absorbing parts of the boundary. The operator $Q[\Psi]$ contains a part resulting from the noise in the direction of motion and a non-linear part that originates from alignment interactions \cite{peruani2008mean}.  The particle density and velocity fields are simply the zeroth and first velocity moments of the field $\Psi$:
\begin{equation}\label{eq:den}
    \rho(\vec{x},t) = \int d\theta \Psi (\vec{x},\theta,t),
\end{equation}
\begin{equation}\label{eq:MFTvel}
    \rho \vec{V} = \int d\theta \vec{v}(\theta)\Psi.
\end{equation}
By integrating the Boltzmann equation over the angles one obtains the mass conservation equation 
\begin{equation}\label{eq:masscons}
    \partial_t \rho(\vec{x},t)  + \nabla_x \cdot \left( \rho(\vec{x},t) \vec{V}(\vec{x},t)\right) =0.
\end{equation}

Since the collision operator is in general non-linear we do not expect a full solution to be available  for $\Psi$ through the method of separation of variables. However, it is instructive to make the somewhat weaker assumption that the two main hydrodynamic fields, after integrating out the direction of motion, are separable, namely $\rho = X(\vec{x}) S(t)$ and $\vec{V} = \vec{u}(\vec{x})f(t)$.  This reduces Eq. (\ref{eq:masscons}) to 

$$- \frac{\dot{S}(t)}{f(t) S(t)} =  \vec{u}(\vec{x})\cdot \nabla_x\log X(\vec{x}) + \nabla_x \cdot \vec{u}(\vec{x}), $$
which is in a separated form, implying immediately that the both sides are equal to some constant which in principle is determined from the boundaries, allowing us to write
\begin{align}
    &\dot S(t) = -\lambda(t) S(t), \label{eq:maineq} \\
    &\lambda(t) = f(t) k,
    \label{eq:MFTrate}
\end{align}
with $k$ a separation constant. This shows that in general we expect the escaping Vicsek particles to behave like a non-stationary Poisson process, with a rate that is some complex function of all the system parameters and time $\lambda(t) = \lambda(D_\theta, r_\text{int},...;t)$.

In the absence of interactions when the non-linear collision term is not present, one can attempt a solution in terms of a fully separated set of variables. Writing the phase-space density as $\Psi = X(\vec{x}) \Theta(\theta) S(t)$ we easily see that the velocity field equation Eq. (\ref{eq:MFTvel}) reduces to 
$$\vec{V} = \int d\theta \vec{v}(\theta)\Theta(\theta),$$
which is time-independent. In this case Eq. (\ref{eq:MFTrate}) becomes a constant, and we expect a stationary Poisson process to be a valid description of the escape process. That the non-interacting system behaves like a stationary Poisson process can also understood from the memorylessness property \cite{ramachandran1979strong}. This property states that, if ones waited some time $t_1$ and no escape has takes place, the probability of having to wait a further time $t_2$ is simply the probability of having to wait a time $t_2$ in the first place. This type of lack of memory, regarding how much time has passed, can be written in terms of the survival probability simply as $S(t_1+t_2) = S(t_1)S(t_2)$ which is only satisfied by an exponential function. We expect the correlations between the particles in the interacting case to break this memoryless property through the fact that the system now depends on its history: flocks of particles may form and escape the system collectively, and the potential size of the clusters is limited by how many particles have already left the system.  We therefore expect the escape rate to be a function of the particle number $\lambda(t) = \lambda[n(t)]$.

Such processes where rates are dependent on density of number of particles are ubiquitous in Nature. In epidemiology, for example, both death and infection rates may have non-trivial density or population size dependencies, which may be traced back to some sort of competition of resources or the simple fact that a higher-density population will have more contacts which act as possible disease transmission routes \cite{yoshida2007global, demirci2011fractional}. These ideas are found in several mathematical models in population ecology, and are typically associated with the Allee effect \cite{dos2015models}. This is the effect where there is a correlation between the general well-being or chance of survival for an individual in a population and that populations size or density.

\begin{figure}[t]
    \centering
    \includegraphics[width = 8.5 cm]{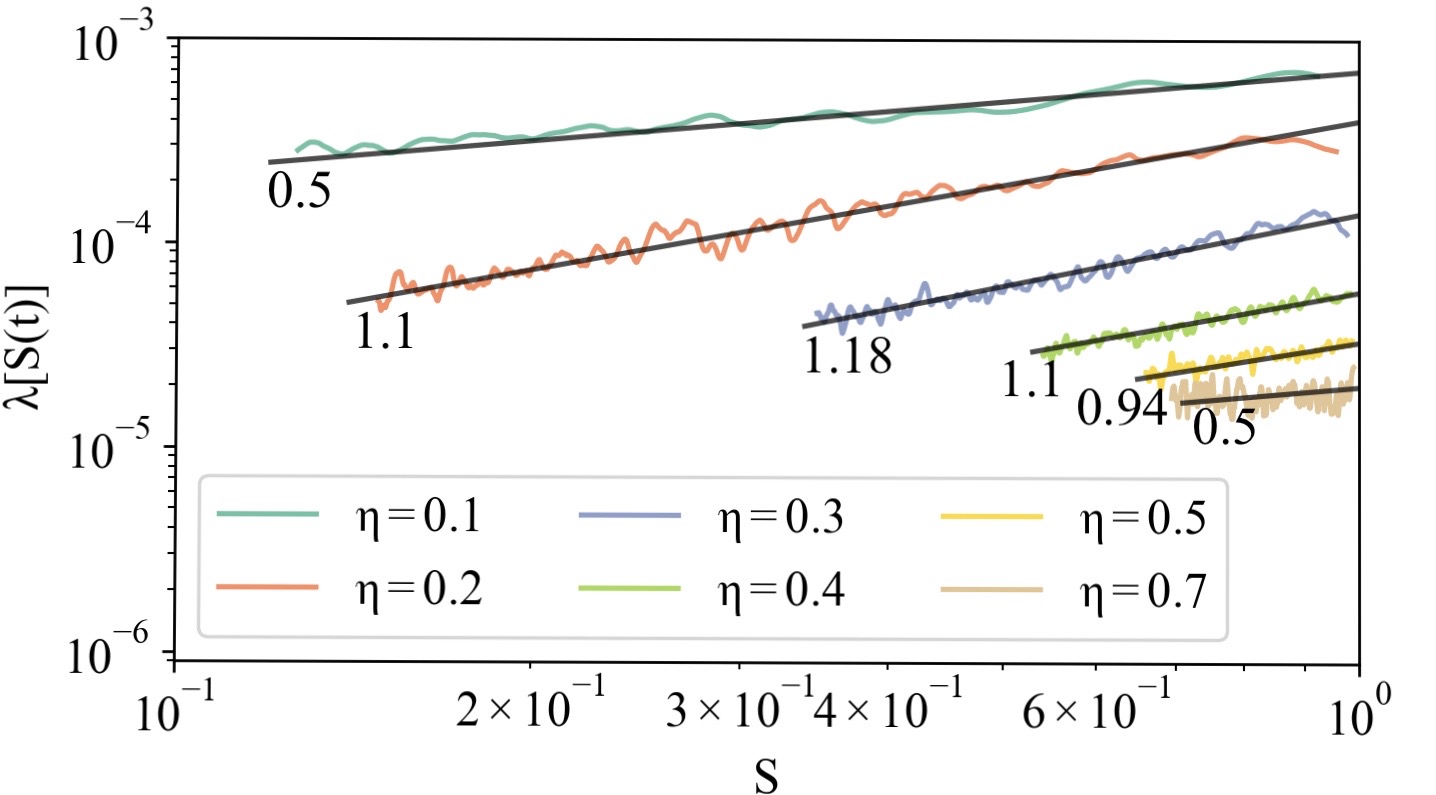}
    \caption{Escape rates plotted as a function of $S(t) = n(t)/n(0)$ showing a power-law behavior. Although the range is insufficient for a highly reliable extraction of the power-law exponent (number attached to each solid black line) the numerical values can be used as initial guesses in a non-linear regression to fit $S(t)$ as a function of time.}
    \label{fig:powerlaw}
\end{figure}

To extract the potential density dependence from the numerical data, we consider Eq. (\ref{eq:maineq}) in the form $\lambda[S] = -\dot{S}/S $, where we used $S = n(t)/n(0)$ to write the rate in terms of the survival probability. Fig. (\ref{fig:powerlaw}) shows these data, where a power-law behavior is observed. This motivates the use of the following power-law ansatz $\lambda[n(t)] = \lambda_0 S
^{\zeta}(t)$, with $S = n/N_0$. This is easily shown to have the solution 
\begin{equation}\label{eq:deformed}
S(t) = \left[1 + \lambda_0 \zeta t\right]^{-1/\zeta}.
\end{equation}
Here the parameter $\lambda_0$ is an escape rate, while the shape parameter $\zeta$ deforms the decaying function $S(t)$ away from the exponential behavior which is regained in the limit $\zeta = 0$. For short times we have an exponential-type behavior $S(t) = 1 - \lambda_0 t+.. $ which is independent of the shape parameter.   For $\zeta>0$, the solution in Eq. (\ref{eq:deformed}) represents a sub-exponential growth at late times, while for negative shape parameters the decay reaches zero at some finite time.

\begin{figure}[ht]
   \centering
    \includegraphics[width = 8.5 cm]{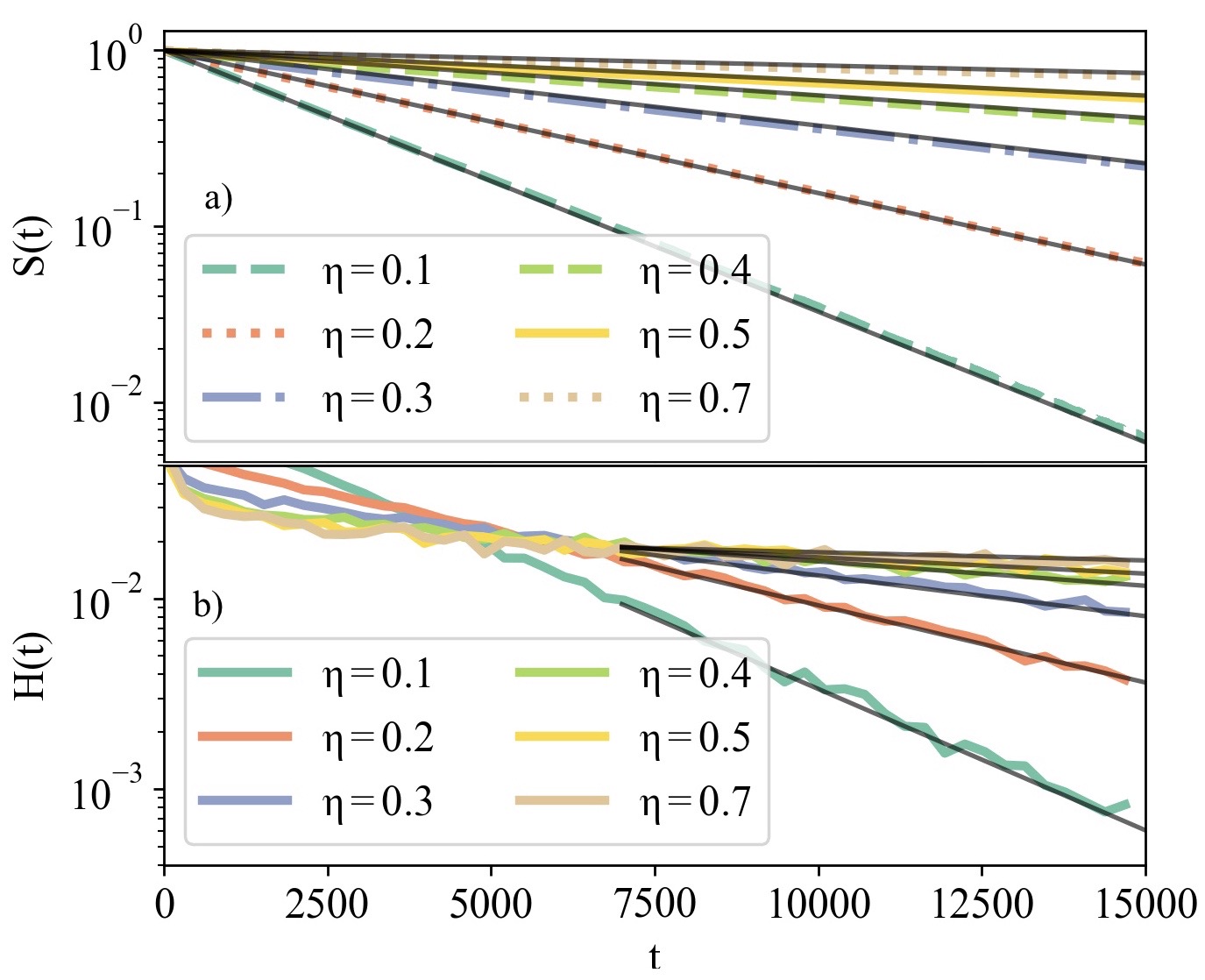}
    \caption{Behavior of the survival probability for self-propelled non-interacting particles. Simulations use $N_0 = 2^9$ particles in a system with radius $R = 100$ in units of the particle step size. a) Survival probability in semi-log plot for some different values of noise strength showing  exponential behavior. b)FHT distribution also consistent with exponential decay.}
    \label{fig:nonint}
\end{figure}

\begin{figure}[ht]
    \centering
    \includegraphics[width = 8.5 cm]{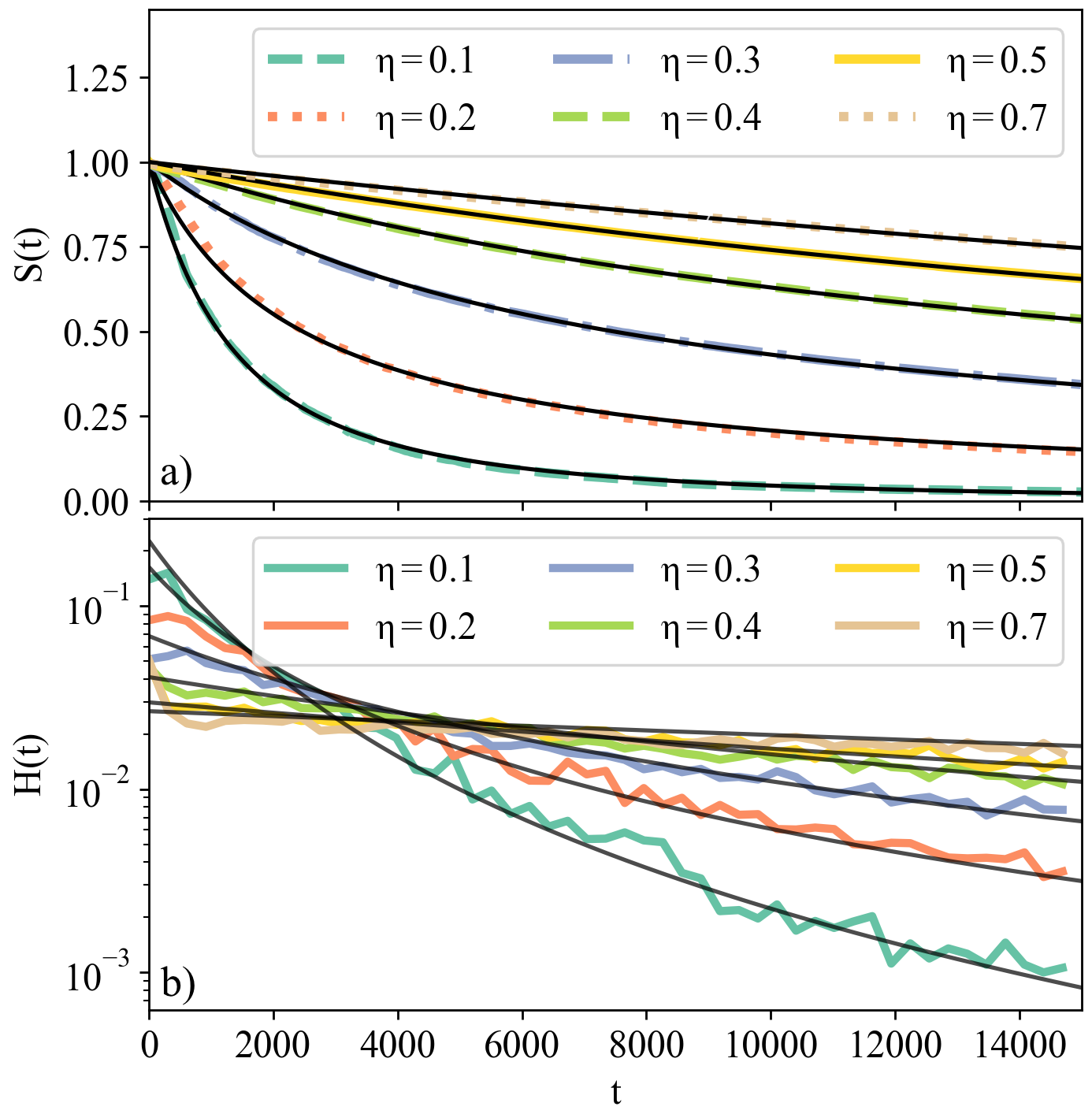}
    \caption{Behavior of the survival probability for self-propelled interacting particles. Simulations use $N_0 = 2^9$ particles in a system with radius $R = 100$ in units of the particle step size. a) Survival probability for some different values of noise strength showing non-exponential behavior together with best fit of Eq. (\ref{eq:deformed}) in solid black lines. The exponent $\zeta$ is taken from Fig. (\ref{fig:powerlaw}). b) FHT distribution also consistent with the same parameters. }
    \label{fig:int}
\end{figure}


\textit{Non-interacting case:} Simulation results from non-interacting self-propelled particles are shown in Fig. (\ref{fig:nonint}), together with best fit exponential lines. We see that both the survival probability and the FHT distribution is exponential at late times as expected, with a rate that decays rapidly as a function of angular noise strength.

\begin{figure}[ht]
    \centering
    \includegraphics[width = 8.5 cm]{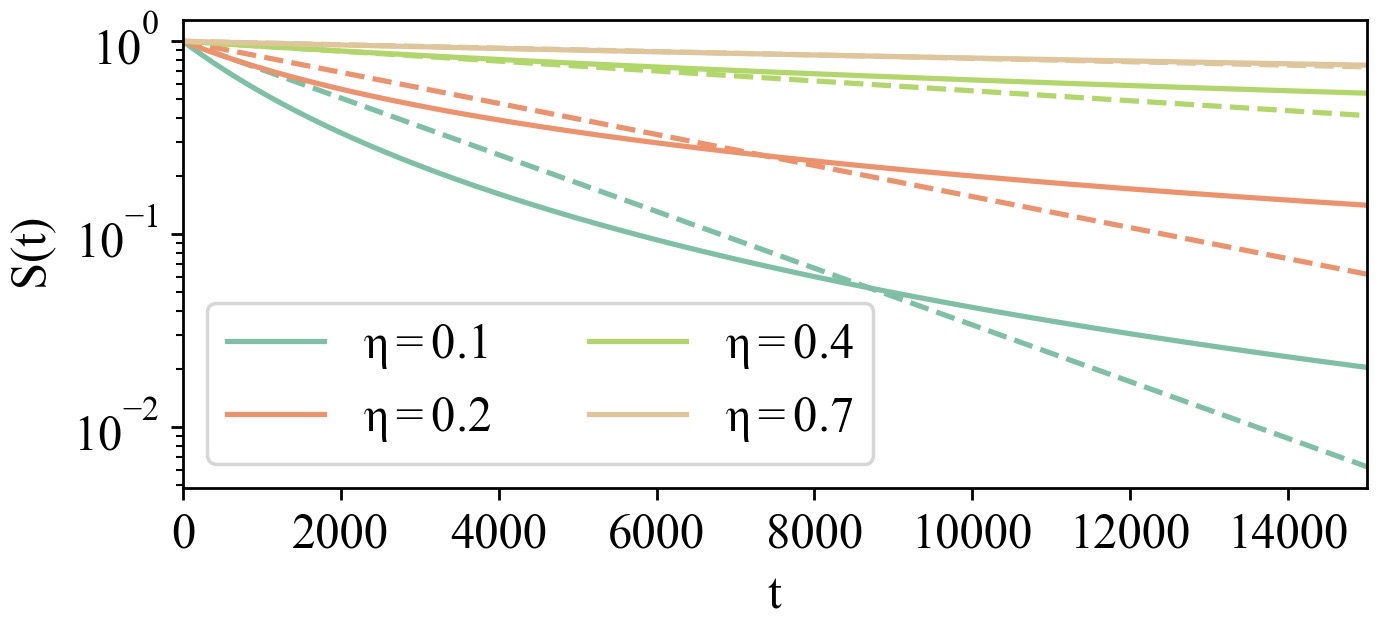}
    \caption{Plot showing the interacting and non-interacting survival probabilities as a function of time for the same system parameters. The interactions make the number of escapes be higher at early times, while the behavior is clearly sub-exponential at late times. }
    \label{fig:cross}
\end{figure}

\textit{Interacting case:} As the alignment interactions in the model are turned on, we expect some deviation from the exponential predictions based on the free theory. Since the particles move collectively in clusters, it is likely that when a single particle finds the absorbing window, so will several other particles in the same cluster. The same may also be said for cases where a cluster misses the window. Results from simulations are shown in Fig. (\ref{fig:int}) together with best fits from the phenomenological model. 

Fig. (\ref{fig:cross}) shows the interacting and non-interacting survival probability together for some chosen values of the noise. We clearly see that while the interacting particles are leaving the system more rapidly at short time scales, they are less efficient at emptying the system at late times. The curves typically cross each other for low noise values while the curves coincide at high noise values. By increasing the initial number of particles, a very similar behavior is expected, except that the crossover time increases since there are more particles present for a longer time to join a flocking state.

\textit{Conclusions:} We have studied the escape problem for self-propelled particles with tunable interaction parameter. In the interacting case, the numerical results agree well with a model where the escape rate is a  power-law in the population fraction, leading to an early time exponential behavior followed by a sub-exponential  decay in time.  Surprisingly, collective alignment effects seem to make the escape process slower in the long run, and faster on short time scales as long as the noise is sufficiently low. In the limit of high noise, fluctuations will dominate over the alignment mechanism and the interacting and non-interacting cases are more or less identical and characterized by exponential decay. \\

\begin{acknowledgements}
The authors thank Gaute Linga, Bjarke F. Nielsen and Vidar Skogvoll for insightful input and discussions during this work. This work was supported by the Research Council of Norway through the Center of Excellence funding scheme, Project No. 262644(PoreLab). L.A. acknowledges support in part by the National Science Foundation under Grant No. NSF PHY-1748958 through the Kavli Institute for Theoretical Physics.  
\end{acknowledgements}

\bibliography{EscapeOlsenEtAl.bib}

\begin{thebibliography}{28}%
\makeatletter
\providecommand \@ifxundefined [1]{%
 \@ifx{#1\undefined}
}%
\providecommand \@ifnum [1]{%
 \ifnum #1\expandafter \@firstoftwo
 \else \expandafter \@secondoftwo
 \fi
}%
\providecommand \@ifx [1]{%
 \ifx #1\expandafter \@firstoftwo
 \else \expandafter \@secondoftwo
 \fi
}%
\providecommand \natexlab [1]{#1}%
\providecommand \enquote  [1]{``#1''}%
\providecommand \bibnamefont  [1]{#1}%
\providecommand \bibfnamefont [1]{#1}%
\providecommand \citenamefont [1]{#1}%
\providecommand \href@noop [0]{\@secondoftwo}%
\providecommand \href [0]{\begingroup \@sanitize@url \@href}%
\providecommand \@href[1]{\@@startlink{#1}\@@href}%
\providecommand \@@href[1]{\endgroup#1\@@endlink}%
\providecommand \@sanitize@url [0]{\catcode `\\12\catcode `\$12\catcode
  `\&12\catcode `\#12\catcode `\^12\catcode `\_12\catcode `\%12\relax}%
\providecommand \@@startlink[1]{}%
\providecommand \@@endlink[0]{}%
\providecommand \url  [0]{\begingroup\@sanitize@url \@url }%
\providecommand \@url [1]{\endgroup\@href {#1}{\urlprefix }}%
\providecommand \urlprefix  [0]{URL }%
\providecommand \Eprint [0]{\href }%
\providecommand \doibase [0]{http://dx.doi.org/}%
\providecommand \selectlanguage [0]{\@gobble}%
\providecommand \bibinfo  [0]{\@secondoftwo}%
\providecommand \bibfield  [0]{\@secondoftwo}%
\providecommand \translation [1]{[#1]}%
\providecommand \BibitemOpen [0]{}%
\providecommand \bibitemStop [0]{}%
\providecommand \bibitemNoStop [0]{.\EOS\space}%
\providecommand \EOS [0]{\spacefactor3000\relax}%
\providecommand \BibitemShut  [1]{\csname bibitem#1\endcsname}%
\let\auto@bib@innerbib\@empty
\bibitem [{\citenamefont {Marchetti}\ \emph {et~al.}(2013)\citenamefont
  {Marchetti}, \citenamefont {Joanny}, \citenamefont {Ramaswamy}, \citenamefont
  {Liverpool}, \citenamefont {Prost}, \citenamefont {Rao},\ and\ \citenamefont
  {Simha}}]{marchetti2013hydrodynamics}%
  \BibitemOpen
  \bibfield  {author} {\bibinfo {author} {\bibfnamefont {M.~C.}\ \bibnamefont
  {Marchetti}}, \bibinfo {author} {\bibfnamefont {J.-F.}\ \bibnamefont
  {Joanny}}, \bibinfo {author} {\bibfnamefont {S.}~\bibnamefont {Ramaswamy}},
  \bibinfo {author} {\bibfnamefont {T.~B.}\ \bibnamefont {Liverpool}}, \bibinfo
  {author} {\bibfnamefont {J.}~\bibnamefont {Prost}}, \bibinfo {author}
  {\bibfnamefont {M.}~\bibnamefont {Rao}}, \ and\ \bibinfo {author}
  {\bibfnamefont {R.~A.}\ \bibnamefont {Simha}},\ }\href@noop {} {\bibfield
  {journal} {\bibinfo  {journal} {Reviews of Modern Physics}\ }\textbf
  {\bibinfo {volume} {85}},\ \bibinfo {pages} {1143} (\bibinfo {year}
  {2013})}\BibitemShut {NoStop}%
\bibitem [{\citenamefont {B{\"a}r}\ \emph {et~al.}(2020)\citenamefont
  {B{\"a}r}, \citenamefont {Gro{\ss}mann}, \citenamefont {Heidenreich},\ and\
  \citenamefont {Peruani}}]{bar2020self}%
  \BibitemOpen
  \bibfield  {author} {\bibinfo {author} {\bibfnamefont {M.}~\bibnamefont
  {B{\"a}r}}, \bibinfo {author} {\bibfnamefont {R.}~\bibnamefont
  {Gro{\ss}mann}}, \bibinfo {author} {\bibfnamefont {S.}~\bibnamefont
  {Heidenreich}}, \ and\ \bibinfo {author} {\bibfnamefont {F.}~\bibnamefont
  {Peruani}},\ }\href@noop {} {\bibfield  {journal} {\bibinfo  {journal}
  {Annual Review of Condensed Matter Physics}\ }\textbf {\bibinfo {volume}
  {11}},\ \bibinfo {pages} {441} (\bibinfo {year} {2020})}\BibitemShut
  {NoStop}%
\bibitem [{\citenamefont {Dombrowski}\ \emph {et~al.}(2004)\citenamefont
  {Dombrowski}, \citenamefont {Cisneros}, \citenamefont {Chatkaew},
  \citenamefont {Goldstein},\ and\ \citenamefont
  {Kessler}}]{dombrowski2004self}%
  \BibitemOpen
  \bibfield  {author} {\bibinfo {author} {\bibfnamefont {C.}~\bibnamefont
  {Dombrowski}}, \bibinfo {author} {\bibfnamefont {L.}~\bibnamefont
  {Cisneros}}, \bibinfo {author} {\bibfnamefont {S.}~\bibnamefont {Chatkaew}},
  \bibinfo {author} {\bibfnamefont {R.~E.}\ \bibnamefont {Goldstein}}, \ and\
  \bibinfo {author} {\bibfnamefont {J.~O.}\ \bibnamefont {Kessler}},\
  }\href@noop {} {\bibfield  {journal} {\bibinfo  {journal} {Physical review
  letters}\ }\textbf {\bibinfo {volume} {93}},\ \bibinfo {pages} {098103}
  (\bibinfo {year} {2004})}\BibitemShut {NoStop}%
\bibitem [{\citenamefont {Riedel}\ \emph {et~al.}(2005)\citenamefont {Riedel},
  \citenamefont {Kruse},\ and\ \citenamefont {Howard}}]{riedel2005self}%
  \BibitemOpen
  \bibfield  {author} {\bibinfo {author} {\bibfnamefont {I.~H.}\ \bibnamefont
  {Riedel}}, \bibinfo {author} {\bibfnamefont {K.}~\bibnamefont {Kruse}}, \
  and\ \bibinfo {author} {\bibfnamefont {J.}~\bibnamefont {Howard}},\
  }\href@noop {} {\bibfield  {journal} {\bibinfo  {journal} {Science}\ }\textbf
  {\bibinfo {volume} {309}},\ \bibinfo {pages} {300} (\bibinfo {year}
  {2005})}\BibitemShut {NoStop}%
\bibitem [{\citenamefont {Walther}\ and\ \citenamefont
  {M{\"u}ller}(2008)}]{walther2008janus}%
  \BibitemOpen
  \bibfield  {author} {\bibinfo {author} {\bibfnamefont {A.}~\bibnamefont
  {Walther}}\ and\ \bibinfo {author} {\bibfnamefont {A.~H.}\ \bibnamefont
  {M{\"u}ller}},\ }\href@noop {} {\bibfield  {journal} {\bibinfo  {journal}
  {Soft Matter}\ }\textbf {\bibinfo {volume} {4}},\ \bibinfo {pages} {663}
  (\bibinfo {year} {2008})}\BibitemShut {NoStop}%
\bibitem [{\citenamefont {Jiang}\ \emph {et~al.}(2010)\citenamefont {Jiang},
  \citenamefont {Yoshinaga},\ and\ \citenamefont {Sano}}]{jiang2010active}%
  \BibitemOpen
  \bibfield  {author} {\bibinfo {author} {\bibfnamefont {H.-R.}\ \bibnamefont
  {Jiang}}, \bibinfo {author} {\bibfnamefont {N.}~\bibnamefont {Yoshinaga}}, \
  and\ \bibinfo {author} {\bibfnamefont {M.}~\bibnamefont {Sano}},\ }\href@noop
  {} {\bibfield  {journal} {\bibinfo  {journal} {Physical review letters}\
  }\textbf {\bibinfo {volume} {105}},\ \bibinfo {pages} {268302} (\bibinfo
  {year} {2010})}\BibitemShut {NoStop}%
\bibitem [{\citenamefont {Bickel}\ \emph {et~al.}(2014)\citenamefont {Bickel},
  \citenamefont {Zecua},\ and\ \citenamefont
  {W\"urger}}]{bickel2014polarization}%
  \BibitemOpen
  \bibfield  {author} {\bibinfo {author} {\bibfnamefont {T.}~\bibnamefont
  {Bickel}}, \bibinfo {author} {\bibfnamefont {G.}~\bibnamefont {Zecua}}, \
  and\ \bibinfo {author} {\bibfnamefont {A.}~\bibnamefont {W\"urger}},\ }\href
  {\doibase 10.1103/PhysRevE.89.050303} {\bibfield  {journal} {\bibinfo
  {journal} {Phys. Rev. E}\ }\textbf {\bibinfo {volume} {89}},\ \bibinfo
  {pages} {050303} (\bibinfo {year} {2014})}\BibitemShut {NoStop}%
\bibitem [{\citenamefont {Kumar}\ \emph {et~al.}(2014)\citenamefont {Kumar},
  \citenamefont {Soni}, \citenamefont {Ramaswamy},\ and\ \citenamefont
  {Sood}}]{kumar2014flocking}%
  \BibitemOpen
  \bibfield  {author} {\bibinfo {author} {\bibfnamefont {N.}~\bibnamefont
  {Kumar}}, \bibinfo {author} {\bibfnamefont {H.}~\bibnamefont {Soni}},
  \bibinfo {author} {\bibfnamefont {S.}~\bibnamefont {Ramaswamy}}, \ and\
  \bibinfo {author} {\bibfnamefont {A.}~\bibnamefont {Sood}},\ }\href@noop {}
  {\bibfield  {journal} {\bibinfo  {journal} {Nature communications}\ }\textbf
  {\bibinfo {volume} {5}},\ \bibinfo {pages} {1} (\bibinfo {year}
  {2014})}\BibitemShut {NoStop}%
\bibitem [{\citenamefont {Kudrolli}\ \emph {et~al.}(2008)\citenamefont
  {Kudrolli}, \citenamefont {Lumay}, \citenamefont {Volfson},\ and\
  \citenamefont {Tsimring}}]{kudrolli2008swarming}%
  \BibitemOpen
  \bibfield  {author} {\bibinfo {author} {\bibfnamefont {A.}~\bibnamefont
  {Kudrolli}}, \bibinfo {author} {\bibfnamefont {G.}~\bibnamefont {Lumay}},
  \bibinfo {author} {\bibfnamefont {D.}~\bibnamefont {Volfson}}, \ and\
  \bibinfo {author} {\bibfnamefont {L.~S.}\ \bibnamefont {Tsimring}},\
  }\href@noop {} {\bibfield  {journal} {\bibinfo  {journal} {Physical review
  letters}\ }\textbf {\bibinfo {volume} {100}},\ \bibinfo {pages} {058001}
  (\bibinfo {year} {2008})}\BibitemShut {NoStop}%
\bibitem [{\citenamefont {Bechinger}\ \emph {et~al.}(2016)\citenamefont
  {Bechinger}, \citenamefont {Di~Leonardo}, \citenamefont {L\"owen},
  \citenamefont {Reichhardt}, \citenamefont {Volpe},\ and\ \citenamefont
  {Volpe}}]{crowdedactive}%
  \BibitemOpen
  \bibfield  {author} {\bibinfo {author} {\bibfnamefont {C.}~\bibnamefont
  {Bechinger}}, \bibinfo {author} {\bibfnamefont {R.}~\bibnamefont
  {Di~Leonardo}}, \bibinfo {author} {\bibfnamefont {H.}~\bibnamefont
  {L\"owen}}, \bibinfo {author} {\bibfnamefont {C.}~\bibnamefont {Reichhardt}},
  \bibinfo {author} {\bibfnamefont {G.}~\bibnamefont {Volpe}}, \ and\ \bibinfo
  {author} {\bibfnamefont {G.}~\bibnamefont {Volpe}},\ }\href {\doibase
  10.1103/RevModPhys.88.045006} {\bibfield  {journal} {\bibinfo  {journal}
  {Rev. Mod. Phys.}\ }\textbf {\bibinfo {volume} {88}},\ \bibinfo {pages}
  {045006} (\bibinfo {year} {2016})}\BibitemShut {NoStop}%
\bibitem [{\citenamefont {Volpe}\ \emph {et~al.}(2014)\citenamefont {Volpe},
  \citenamefont {Gigan},\ and\ \citenamefont {Volpe}}]{volpe2014simulation}%
  \BibitemOpen
  \bibfield  {author} {\bibinfo {author} {\bibfnamefont {G.}~\bibnamefont
  {Volpe}}, \bibinfo {author} {\bibfnamefont {S.}~\bibnamefont {Gigan}}, \ and\
  \bibinfo {author} {\bibfnamefont {G.}~\bibnamefont {Volpe}},\ }\href@noop {}
  {\bibfield  {journal} {\bibinfo  {journal} {American Journal of Physics}\
  }\textbf {\bibinfo {volume} {82}},\ \bibinfo {pages} {659} (\bibinfo {year}
  {2014})}\BibitemShut {NoStop}%
\bibitem [{\citenamefont {Yang}\ \emph {et~al.}(2014)\citenamefont {Yang},
  \citenamefont {Manning},\ and\ \citenamefont
  {Marchetti}}]{yang2014aggregation}%
  \BibitemOpen
  \bibfield  {author} {\bibinfo {author} {\bibfnamefont {X.}~\bibnamefont
  {Yang}}, \bibinfo {author} {\bibfnamefont {M.~L.}\ \bibnamefont {Manning}}, \
  and\ \bibinfo {author} {\bibfnamefont {M.~C.}\ \bibnamefont {Marchetti}},\
  }\href@noop {} {\bibfield  {journal} {\bibinfo  {journal} {Soft matter}\
  }\textbf {\bibinfo {volume} {10}},\ \bibinfo {pages} {6477} (\bibinfo {year}
  {2014})}\BibitemShut {NoStop}%
\bibitem [{\citenamefont {Mijalkov}\ and\ \citenamefont
  {Volpe}(2013)}]{mijalkov2013sorting}%
  \BibitemOpen
  \bibfield  {author} {\bibinfo {author} {\bibfnamefont {M.}~\bibnamefont
  {Mijalkov}}\ and\ \bibinfo {author} {\bibfnamefont {G.}~\bibnamefont
  {Volpe}},\ }\href@noop {} {\bibfield  {journal} {\bibinfo  {journal} {Soft
  Matter}\ }\textbf {\bibinfo {volume} {9}},\ \bibinfo {pages} {6376} (\bibinfo
  {year} {2013})}\BibitemShut {NoStop}%
\bibitem [{\citenamefont {Kumar}\ \emph {et~al.}(2019)\citenamefont {Kumar},
  \citenamefont {Gupta}, \citenamefont {Soni}, \citenamefont {Ramaswamy},\ and\
  \citenamefont {Sood}}]{kumar2019trapping}%
  \BibitemOpen
  \bibfield  {author} {\bibinfo {author} {\bibfnamefont {N.}~\bibnamefont
  {Kumar}}, \bibinfo {author} {\bibfnamefont {R.~K.}\ \bibnamefont {Gupta}},
  \bibinfo {author} {\bibfnamefont {H.}~\bibnamefont {Soni}}, \bibinfo {author}
  {\bibfnamefont {S.}~\bibnamefont {Ramaswamy}}, \ and\ \bibinfo {author}
  {\bibfnamefont {A.}~\bibnamefont {Sood}},\ }\href@noop {} {\bibfield
  {journal} {\bibinfo  {journal} {Physical Review E}\ }\textbf {\bibinfo
  {volume} {99}},\ \bibinfo {pages} {032605} (\bibinfo {year}
  {2019})}\BibitemShut {NoStop}%
\bibitem [{\citenamefont {Palagi}\ and\ \citenamefont
  {Fischer}(2018)}]{palagi2018bioinspired}%
  \BibitemOpen
  \bibfield  {author} {\bibinfo {author} {\bibfnamefont {S.}~\bibnamefont
  {Palagi}}\ and\ \bibinfo {author} {\bibfnamefont {P.}~\bibnamefont
  {Fischer}},\ }\href@noop {} {\bibfield  {journal} {\bibinfo  {journal}
  {Nature Reviews Materials}\ }\textbf {\bibinfo {volume} {3}},\ \bibinfo
  {pages} {113} (\bibinfo {year} {2018})}\BibitemShut {NoStop}%
\bibitem [{\citenamefont {Gompper}\ \emph {et~al.}(2020)\citenamefont
  {Gompper}, \citenamefont {Winkler}, \citenamefont {Speck}, \citenamefont
  {Solon}, \citenamefont {Nardini}, \citenamefont {Peruani}, \citenamefont
  {L{\"o}wen}, \citenamefont {Golestanian}, \citenamefont {Kaupp},
  \citenamefont {Alvarez} \emph {et~al.}}]{gompper20202020}%
  \BibitemOpen
  \bibfield  {author} {\bibinfo {author} {\bibfnamefont {G.}~\bibnamefont
  {Gompper}}, \bibinfo {author} {\bibfnamefont {R.~G.}\ \bibnamefont
  {Winkler}}, \bibinfo {author} {\bibfnamefont {T.}~\bibnamefont {Speck}},
  \bibinfo {author} {\bibfnamefont {A.}~\bibnamefont {Solon}}, \bibinfo
  {author} {\bibfnamefont {C.}~\bibnamefont {Nardini}}, \bibinfo {author}
  {\bibfnamefont {F.}~\bibnamefont {Peruani}}, \bibinfo {author} {\bibfnamefont
  {H.}~\bibnamefont {L{\"o}wen}}, \bibinfo {author} {\bibfnamefont
  {R.}~\bibnamefont {Golestanian}}, \bibinfo {author} {\bibfnamefont {U.~B.}\
  \bibnamefont {Kaupp}}, \bibinfo {author} {\bibfnamefont {L.}~\bibnamefont
  {Alvarez}},  \emph {et~al.},\ }\href@noop {} {\bibfield  {journal} {\bibinfo
  {journal} {Journal of Physics: Condensed Matter}\ }\textbf {\bibinfo {volume}
  {32}},\ \bibinfo {pages} {193001} (\bibinfo {year} {2020})}\BibitemShut
  {NoStop}%
\bibitem [{\citenamefont {Schuss}\ \emph {et~al.}(2007)\citenamefont {Schuss},
  \citenamefont {Singer},\ and\ \citenamefont {Holcman}}]{schuss2007narrow}%
  \BibitemOpen
  \bibfield  {author} {\bibinfo {author} {\bibfnamefont {Z.}~\bibnamefont
  {Schuss}}, \bibinfo {author} {\bibfnamefont {A.}~\bibnamefont {Singer}}, \
  and\ \bibinfo {author} {\bibfnamefont {D.}~\bibnamefont {Holcman}},\
  }\href@noop {} {\bibfield  {journal} {\bibinfo  {journal} {Proceedings of the
  National Academy of Sciences}\ }\textbf {\bibinfo {volume} {104}},\ \bibinfo
  {pages} {16098} (\bibinfo {year} {2007})}\BibitemShut {NoStop}%
\bibitem [{\citenamefont {Singer}\ \emph {et~al.}(2008)\citenamefont {Singer},
  \citenamefont {Schuss},\ and\ \citenamefont {Holcman}}]{singer2008narrow}%
  \BibitemOpen
  \bibfield  {author} {\bibinfo {author} {\bibfnamefont {A.}~\bibnamefont
  {Singer}}, \bibinfo {author} {\bibfnamefont {Z.}~\bibnamefont {Schuss}}, \
  and\ \bibinfo {author} {\bibfnamefont {D.}~\bibnamefont {Holcman}},\
  }\href@noop {} {\bibfield  {journal} {\bibinfo  {journal} {Physical Review
  E}\ }\textbf {\bibinfo {volume} {78}},\ \bibinfo {pages} {051111} (\bibinfo
  {year} {2008})}\BibitemShut {NoStop}%
\bibitem [{\citenamefont {Bauer}\ and\ \citenamefont
  {Bertsch}(1990)}]{bauer1990decay}%
  \BibitemOpen
  \bibfield  {author} {\bibinfo {author} {\bibfnamefont {W.}~\bibnamefont
  {Bauer}}\ and\ \bibinfo {author} {\bibfnamefont {G.}~\bibnamefont
  {Bertsch}},\ }\href@noop {} {\bibfield  {journal} {\bibinfo  {journal}
  {Physical review letters}\ }\textbf {\bibinfo {volume} {65}},\ \bibinfo
  {pages} {2213} (\bibinfo {year} {1990})}\BibitemShut {NoStop}%
\bibitem [{\citenamefont {Mori}\ \emph {et~al.}(2020)\citenamefont {Mori},
  \citenamefont {Le~Doussal}, \citenamefont {Majumdar},\ and\ \citenamefont
  {Schehr}}]{mori2020universal}%
  \BibitemOpen
  \bibfield  {author} {\bibinfo {author} {\bibfnamefont {F.}~\bibnamefont
  {Mori}}, \bibinfo {author} {\bibfnamefont {P.}~\bibnamefont {Le~Doussal}},
  \bibinfo {author} {\bibfnamefont {S.~N.}\ \bibnamefont {Majumdar}}, \ and\
  \bibinfo {author} {\bibfnamefont {G.}~\bibnamefont {Schehr}},\ }\href@noop {}
  {\bibfield  {journal} {\bibinfo  {journal} {Physical Review Letters}\
  }\textbf {\bibinfo {volume} {124}},\ \bibinfo {pages} {090603} (\bibinfo
  {year} {2020})}\BibitemShut {NoStop}%
\bibitem [{\citenamefont {Vicsek}\ \emph {et~al.}(1995)\citenamefont {Vicsek},
  \citenamefont {Czir{\'o}k}, \citenamefont {Ben-Jacob}, \citenamefont
  {Cohen},\ and\ \citenamefont {Shochet}}]{vicsek1995novel}%
  \BibitemOpen
  \bibfield  {author} {\bibinfo {author} {\bibfnamefont {T.}~\bibnamefont
  {Vicsek}}, \bibinfo {author} {\bibfnamefont {A.}~\bibnamefont {Czir{\'o}k}},
  \bibinfo {author} {\bibfnamefont {E.}~\bibnamefont {Ben-Jacob}}, \bibinfo
  {author} {\bibfnamefont {I.}~\bibnamefont {Cohen}}, \ and\ \bibinfo {author}
  {\bibfnamefont {O.}~\bibnamefont {Shochet}},\ }\href@noop {} {\bibfield
  {journal} {\bibinfo  {journal} {Physical review letters}\ }\textbf {\bibinfo
  {volume} {75}},\ \bibinfo {pages} {1226} (\bibinfo {year}
  {1995})}\BibitemShut {NoStop}%
\bibitem [{\citenamefont {Ginelli}(2016)}]{ginelli2016physics}%
  \BibitemOpen
  \bibfield  {author} {\bibinfo {author} {\bibfnamefont {F.}~\bibnamefont
  {Ginelli}},\ }\href@noop {} {\bibfield  {journal} {\bibinfo  {journal} {The
  European Physical Journal Special Topics}\ }\textbf {\bibinfo {volume}
  {225}},\ \bibinfo {pages} {2099} (\bibinfo {year} {2016})}\BibitemShut
  {NoStop}%
\bibitem [{\citenamefont {Chat{\'e}}\ \emph {et~al.}(2008)\citenamefont
  {Chat{\'e}}, \citenamefont {Ginelli}, \citenamefont {Gr{\'e}goire},
  \citenamefont {Peruani},\ and\ \citenamefont {Raynaud}}]{chate2008modeling}%
  \BibitemOpen
  \bibfield  {author} {\bibinfo {author} {\bibfnamefont {H.}~\bibnamefont
  {Chat{\'e}}}, \bibinfo {author} {\bibfnamefont {F.}~\bibnamefont {Ginelli}},
  \bibinfo {author} {\bibfnamefont {G.}~\bibnamefont {Gr{\'e}goire}}, \bibinfo
  {author} {\bibfnamefont {F.}~\bibnamefont {Peruani}}, \ and\ \bibinfo
  {author} {\bibfnamefont {F.}~\bibnamefont {Raynaud}},\ }\href@noop {}
  {\bibfield  {journal} {\bibinfo  {journal} {The European Physical Journal B}\
  }\textbf {\bibinfo {volume} {64}},\ \bibinfo {pages} {451} (\bibinfo {year}
  {2008})}\BibitemShut {NoStop}%
\bibitem [{\citenamefont {Peruani}\ \emph {et~al.}(2008)\citenamefont
  {Peruani}, \citenamefont {Deutsch},\ and\ \citenamefont
  {B{\"a}r}}]{peruani2008mean}%
  \BibitemOpen
  \bibfield  {author} {\bibinfo {author} {\bibfnamefont {F.}~\bibnamefont
  {Peruani}}, \bibinfo {author} {\bibfnamefont {A.}~\bibnamefont {Deutsch}}, \
  and\ \bibinfo {author} {\bibfnamefont {M.}~\bibnamefont {B{\"a}r}},\
  }\href@noop {} {\bibfield  {journal} {\bibinfo  {journal} {The European
  Physical Journal Special Topics}\ }\textbf {\bibinfo {volume} {157}},\
  \bibinfo {pages} {111} (\bibinfo {year} {2008})}\BibitemShut {NoStop}%
\bibitem [{\citenamefont {Ramachandran}(1979)}]{ramachandran1979strong}%
  \BibitemOpen
  \bibfield  {author} {\bibinfo {author} {\bibfnamefont {B.}~\bibnamefont
  {Ramachandran}},\ }\href@noop {} {\bibfield  {journal} {\bibinfo  {journal}
  {Sankhy{\=a}: The Indian Journal of Statistics, Series A}\ ,\ \bibinfo
  {pages} {244}} (\bibinfo {year} {1979})}\BibitemShut {NoStop}%
\bibitem [{\citenamefont {Yoshida}\ and\ \citenamefont
  {Hara}(2007)}]{yoshida2007global}%
  \BibitemOpen
  \bibfield  {author} {\bibinfo {author} {\bibfnamefont {N.}~\bibnamefont
  {Yoshida}}\ and\ \bibinfo {author} {\bibfnamefont {T.}~\bibnamefont {Hara}},\
  }\href@noop {} {\bibfield  {journal} {\bibinfo  {journal} {Journal of
  Computational and Applied Mathematics}\ }\textbf {\bibinfo {volume} {201}},\
  \bibinfo {pages} {339} (\bibinfo {year} {2007})}\BibitemShut {NoStop}%
\bibitem [{\citenamefont {Demirci}\ \emph {et~al.}(2011)\citenamefont
  {Demirci}, \citenamefont {Unal},\ and\ \citenamefont
  {Ozalp}}]{demirci2011fractional}%
  \BibitemOpen
  \bibfield  {author} {\bibinfo {author} {\bibfnamefont {E.}~\bibnamefont
  {Demirci}}, \bibinfo {author} {\bibfnamefont {A.}~\bibnamefont {Unal}}, \
  and\ \bibinfo {author} {\bibfnamefont {N.}~\bibnamefont {Ozalp}},\
  }\href@noop {} {\bibfield  {journal} {\bibinfo  {journal} {Hacettepe Journal
  of Mathematics and Statistics}\ }\textbf {\bibinfo {volume} {40}},\ \bibinfo
  {pages} {287} (\bibinfo {year} {2011})}\BibitemShut {NoStop}%
\bibitem [{\citenamefont {dos Santos}\ \emph {et~al.}(2015)\citenamefont {dos
  Santos}, \citenamefont {Ribeiro},\ and\ \citenamefont
  {Martinez}}]{dos2015models}%
  \BibitemOpen
  \bibfield  {author} {\bibinfo {author} {\bibfnamefont {R.~V.}\ \bibnamefont
  {dos Santos}}, \bibinfo {author} {\bibfnamefont {F.~L.}\ \bibnamefont
  {Ribeiro}}, \ and\ \bibinfo {author} {\bibfnamefont {A.~S.}\ \bibnamefont
  {Martinez}},\ }\href@noop {} {\bibfield  {journal} {\bibinfo  {journal}
  {Journal of theoretical biology}\ }\textbf {\bibinfo {volume} {385}},\
  \bibinfo {pages} {143} (\bibinfo {year} {2015})}\BibitemShut {NoStop}%
\end{thebibliography}%


\end{document}